\documentclass[12pt]{article}
\usepackage{amsmath}
\usepackage{amsfonts}
\usepackage{amssymb}
\usepackage{graphicx}
\begin{document}

\title{\bf Dynamics of  an Anisotropic Universe in $f(R,T)$ Theory}

\author{B. Mishra \footnote{Department of Mathematics, Birla Institute of Technology and
Science-Pilani, Hyderabad Campus, Hyderabad-500078, India, E-mail:bivudutta@yahoo.com}, Sankarsan Tarai\footnote{Department of Mathematics, Birla Institute of Technology and
Science-Pilani, Hyderabad Campus, Hyderabad-500078, India, E-mail:tsankarsan87@gmail.com} and S. K. Tripathy\footnote{Department of Physics, Indira Gandhi Institute of Technology, Sarang, Dhenkanal, Odisha-759146, India, E-mail:tripathy\_ sunil@rediffmail.com}
}

\maketitle

\begin{abstract}
Dynamics of an anisotropic universe is studied in $f(R,T)$ gravity using a rescaled functional $f(R,T)$. Three models have been constructed assuming a power law expansion of the universe. Physical features of the models are discussed. The model parameters are constrained from a dimensional analysis. It is found from the work that, the $BVI_h$ model in the modified gravity can favour quintessence and phantom phase. 
\end{abstract}


\section{Introduction} 

In the past one decade the idea of modifying gravity on cosmological scale has attracted a lot of attention. It gains momentum through the theoretical developments involving higher dimensional theories and in constructing renormalizable theories of gravity. Modified gravity represents an intriguing possibility for resolving the theoretical challenge posed by late time acceleration. However, it turns out to be extremely difficult to modify general relativity at low energy regime without violating  observational constraints.

Observations from type Ia supernova \cite{Reiss98,Perlm99, Bahcall99},  large scale structures \cite{Tegmark04, Abaz04, Pope04} and cosmic microwave background \cite{Bennet03, Spergel03} confirm that the cosmic expansion is accelerating. The very cause behind this acceleration may be a mysterious energy source called dark energy which accounts for two-third of the total energy budget of the universe. 

The role of modified theories to understand the mechanism behind the cosmic speed up is more promising.
The objective behind such theories is to explain the puzzling late time cosmic dynamics without the need of dark energy components in the field equations. Several modified theories of gravity have been proposed in recent times. Of these a few models such as $f(R)$ gravity \cite{Nojiri06, Nojiri07}, $f(T)$ gravity \cite{Linder10, Myrz11, Chen11,Dent11, Harko14} and $f(G)$ gravity \cite{Nojiri05, Li07, Kofinas14} are studied widely. In $f(R)$ theory, a more general function of the Ricci Scalar $R=g^{\mu \nu} R_{\mu \nu}$ is used in place of $R$ whereas $f(T)$ gravity is a generalized version of teleparallel gravity. Motivated by the success of cosmological constant as a simple and good candidate of dark energy, some matter field is also coupled with the function of $R$ in the geometry side of the action in some modified theories of gravity ($f(R,\mathcal{L}_m$) theory). 

Along the line of interest of incorporating some matter components in the action geometry, $f(R,T)$ theory has been proposed by Harko et al. \cite{Harko11} which, of late, has been an interesting framework to investigate accelerating models. Moreover, the reconstruction of arbitrary FRW cosmologies is possible by an appropriate choice of the functional $f(R,T)$.  Many authors have investigated the astrophysical and cosmological implications of the $f(R,T)$ gravity \cite{Myrz12, Sharif12, Houndjo12, Alva13}. Jamil et al.\cite{Jamil12} have reconstructed some cosmological models for some specific forms of $f(R,T)$ in this modified gravity. Shamir et al. \cite{Shamir12} obtained exact solution of anisotropic Bianchi type-I and type-V cosmological models  whereas Chaubey and Shukla \cite{Chau13} have obtained a new class of Bianchi cosmological models using special law of variation of parameter. Using a decoupled form of $f(R,T)$ i.e. $f(R,T)= f(R) +f(T)$ for Bianchi type $V $ universe, Ahmed and Pradhan \cite{Ahmed14} have studied the energy conditions of perfect fluid cosmological models and Yadav \cite{Yadav13} obtained some string solutions.

Singh and Kumar \cite{Singh14} have investigated the effect of bulk viscosity in $f(R,T)$ theory and suggested that inclusion of dissipative energy sources like bulk viscosity may be able to explain the early and late time accelerations of the universe. Mishra and Sahoo \cite{Mishra14} have studied Bianchi type $VI_h$ cosmological models assuming $f(R,T)=R+2f(T)$. Samanta \cite{Samanta13} has obtained exact solution of $f(R,T)$ gravitational field equations in Kantowski-Sachs space time and Shamir \cite{Shamir15} has constructed some cosmological models in Bianchi type V space-time.  In the frame work of this modified gravity, recently, Mishra et al. \cite{Mishra15} have presented the Einstein-Rosen non-static cosmological model with quadratic form of $f (R,T )$ gravity. Sahoo et al. \cite{Sahoo16} have investigated the background cosmology of power law and exponential law of volumetric expansion in $f(R,T)$ gravity in Kaluza-Klein model. 

In this work, we have constructed some cosmological models in $f(R,T)$ theory for general Bianchi type $VI_h$ ($BVI_h$)Universe. Our work is organised as follows. In section 2, the basic formalism of the $f(R,T) $ theory has been presented. The dynamics of anisotropic $VI_h$ model has been presented in a general form.   In section 3, some specific models have been constructed to study the dynamics. We presented the conclusions of the works at the end in section-4.

\section{Basic Formalism}

The modified four dimensional Einstein-Hilbert action in $f(R,T)$ gravity theory with a specific choice of matter Lagrangian can be considered as 

\begin{equation}
S=\frac{1}{16\pi} \int f(R,T)\sqrt{-g} d^4x-\int p \sqrt{-g}d^4x.\label{eq:1}
\end{equation} 
Here, $f(R,T)$ is an arbitrary function of $R$ and  $T$. The universe is considered to be filled with a perfect fluid with pressure $p$. The trace $T=g^{ij}T_{ij}$ of the energy-momentum  tensor  is obtained from $T_{ij}=(\rho+p)u_i u_j-p g_{ij}$. $\rho$ is the rest energy density and $u^i=\delta_0^i$ is the four velocity vector.

The field equations can be obtained from the action in \eqref{eq:1} for the choice of the functional $f(R,T)=f(R)+f(T)$ as \cite{Harko11},

\begin{eqnarray} \label{eq:2}
f_R R_{ij}-\frac{1}{2}f(R)g_{ij}+\left(g_{ij}\Box-\nabla_i \nabla_j\right)f_R(R) &=& 8\pi T_{ij}+f_T(T)T_{ij},\nonumber\\ 
&+& \left[pf_T(T)+\frac{1}{2}f(T) \right] g_{ij}.
\end{eqnarray}
$f_R=\frac{\partial f(R)}{\partial R}$ and $f_T=\frac{\partial f(T)}{\partial T}$ are the partial differentiation of the respective functional with respect to their arguments. The functional $f(R,T)$ can be chosen arbitrarily to get viable cosmological models. Here we chose the functional $f(R)$ and $f(T)$ to be linear in their arguments: $f(R)=\lambda R$ and $f(T)=\lambda T$, so that $f(R,T)= \lambda (R+T)$. $\lambda$ is a constant scaling factor. 

Eq. \eqref{eq:2} now reduces to

\begin{equation}
\label{eq:3}
R_{ij}-\frac{1}{2}Rg_{ij}=\left(\frac{8\pi+\lambda}{\lambda}\right)T_{ij}+\Lambda (T) g_{ij}.
\end{equation}
$\Lambda (T)=p+\frac{1}{2}T$ can now be identified with the cosmological constant which instead of being a pure constant evolves with cosmic time. 

For a spatially homogeneous and anisotropic Bianchi type $VI_{h}$ ($BVI_h$) space time considered in the form
\begin{equation} \label{eq:4}
ds^2 = dt^2 - A^2dx^2- B^2e^{2x}dy^2 - C^2e^{2hx}dz^2,
\end{equation}
eq.\eqref{eq:3} can be explicitly written as

\begin{equation} \label{eq:5}
\frac{\ddot{B}}{B}+\frac{\ddot{C}}{C}+\frac{\dot{B}\dot{C}}{BC}- \frac{h}{A^2}= \left( \frac{16\pi+3\lambda}{2\lambda}\right)p -\frac{\rho}{2}   
\end{equation}
\begin{equation} \label{eq:6}
\frac{\ddot{A}}{A}+\frac{\ddot{C}}{C}+\frac{\dot{A}\dot{C}}{AC}- \frac{h^2}{A^2}=\left( \frac{16\pi+3\lambda}{2\lambda}\right)p -\frac{\rho}{2}  
\end{equation}
\begin{equation} \label{eq:7}
\frac{\ddot{A}}{A}+\frac{\ddot{B}}{B}+\frac{\dot{A}\dot{B}}{AB}- \frac{1}{A^2}=\left( \frac{16\pi+3\lambda}{2\lambda}\right)p -\frac{\rho}{2}   
\end{equation}
\begin{equation} \label{eq:8}
-\frac{\dot{A}\dot{B}}{AB}-\frac{\dot{B}\dot{C}}{BC}-\frac{\dot{C}\dot{A}}{CA}+\frac{1+h+h^2}{A^2}=
\left( \frac{16\pi+3\lambda}{2\lambda}\right)\rho -\frac{p}{2}   
\end{equation}
\begin{equation} \label{eq:9}
\frac{\dot{B}}{B}+ h\frac{\dot{C}}{C}- (1+h)\frac{\dot{A}}{A}=0
\end{equation}
where the metric potentials $A$, $B$ and $C$ are functions of cosmic time $t$. The constant exponent $h$ decides the behaviour of the model and can take integral values such as $-1,\ 0$ and 1. An over dot on a field variable denotes differentiation with respect to time $t$. We define the directional Hubble parameters along different directions as $H_{x}=\frac{\dot{A}}{A}$, $H_{y}=\frac{\dot{B}}{B}$ and $H_{z}=\frac{\dot{C}}{C}$. The mean Hubble parameter becomes $H=\frac{1}{3}(H_{x}+H_{y}+H_{z})$. The field equations \eqref{eq:5}- \eqref{eq:9} can now be expressed as
\begin{equation}\label{eq:10}
\dot{H_{y}}+\dot{H_{z}}+H^{2}_{y}+H^{2}_{z}+H_{y}H_{z}-\frac{h}{A^2}=\alpha p-\frac{\rho}{2},
\end{equation}
\begin{equation}\label{eq:11}
\dot{H_{x}}+\dot{H_{z}}+H^{2}_{x}+H^{2}_{z}+H_{x}H_{z}-\frac{h^2}{A^2}=\alpha p-\frac{\rho}{2},
\end{equation}
\begin{equation}\label{eq:12}
\dot{H_{x}}+\dot{H_{y}}+H^{2}_{x}+H^{2}_{y}+H_{x}H_{y}-\frac{1}{A^2}=\alpha p-\frac{\rho}{2},
\end{equation}
\begin{equation}\label{eq:13}
-H_{x}H_{y}-H_{y}H_{z}-H_{z}H_{x}+\frac{1+h+h^2}{A^2}=\alpha \rho-\frac{p}{2},
\end{equation}
\begin{equation}\label{eq:14}
H_{y}+hH_{z}-(1+h)H_{x}=0,
\end{equation}
where $\alpha=\left(\frac{16 \pi+3 \lambda}{2 \lambda}\right)$. 

The pressure $p$ and rest energy density $\rho$ can be obtained from  \eqref{eq:12}- \eqref{eq:13} in general forms as

\begin{eqnarray}
p &=& \frac{2}{(4\alpha^{2}-1)}\left[2\alpha \chi(H_x,H_y)-\xi(H_x,H_y,H_z,h)\right],\label{eq:15}\\
\rho &=& \frac{2}{(4\alpha^{2}-1)}\left[\chi(H_x,H_y)-2\alpha\xi(H_x,H_y,H_z,h)\right],\label{eq:16}
\end{eqnarray}
where $\chi(H_x,H_y)=\dot{H_{x}}+\dot{H_{y}}+H^{2}_{x}+H^{2}_{y}+H_{x} H_{y}-\frac{1}{A^2}$ and $\xi(H_x,H_y,H_z,h)=H_{x}H_{y}+H_{y}H_{z}+H_{z}H_{x}-\frac{1+h+h^2}{A^2}$.

From eqs. \eqref{eq:15} and \eqref{eq:16}, we obtain the equation of state parameter $\omega=\frac{p}{\rho}$ and the effective cosmological constant $\Lambda$ as
\begin{eqnarray}
\omega &=& 2\alpha+\frac{(4\alpha^2-1)\xi(H_x,H_y,h)}{\chi(H_x,H_y)-2\alpha\xi(H_x,H_y,H_z,h)},\label{eq:17}\\
\Lambda & = & -\frac{\chi(H_x,H_y)+\xi(H_x,H_y,H_z,h)}{(2\alpha+1)}.\label{eq:18}
\end{eqnarray}

The equations \eqref{eq:15}-\eqref{eq:18} provide the dynamical behaviour of the universe. However, the dynamics can only be assessed if the behaviour of these properties are known in terms of the directional Hubble rates for a given value of the exponent $h$. In other words, the formalism as described above can help us to study a background cosmology for an assumed dynamics of the universe. In this context, we can consider the power law cosmology, where the cosmic expansion is governed through a volume scale factor of the form $v=t^m$. $m$ is an arbitrary positive constant usually determined from the background cosmology. Power law cosmology has been widely studied in recent times because of its functional simplicity and ability to provide a first hand information about the dynamics of the universe. For such an assumption, the radius scale factor can be $a=(ABC)^{\frac{1}{3}}=t^{\frac{m}{3}}$. The deceleration parameter for power law expansion of the universe is a constant quantity: $q=-1+\frac{3}{m}$, which can be negative for  $m>3$ and positive for $m<3$. It is worth to mention here that, a positive $q$ describes a decelerating universe whereas a negative $q$ describes an accelerating universe. In order to keep a pace with the recent observational data favouring an accelerating universe, $m$ should be greater than 3.

Some other kinematical parameters of the universe are the scalar expansion $\theta$, shear scalar $\sigma^2$ and the average anisotropy parameter $\mathcal{A}$ defined respectively as
\begin{eqnarray}
\theta &=& \Sigma H_i\label{eq:19}\\
\sigma^2 &=& \frac{1}{2}\sigma_{ij}\sigma^{ij}=\frac{1}{2}\biggl(\Sigma H_i^2-\frac{1}{3}\theta^2\biggr),\label{eq:20}\\
\mathcal{A} &=& \frac{1}{3}\Sigma \left(\frac{\Delta H_i}{H}\right)^2,\label{eq:21}
\end{eqnarray}
where $\Delta H_i=H_i-H$ with $ i=x,y,z$. $\mathcal{A}$ is a measure of deviation from isotropic expansion. One can get the isotropic behaviour of the model for $\mathcal{A}=0$. For the power law cosmology with $v=t^m$, the scalar expansion becomes $\theta=\frac{m}{t}$ and consequently the mean Hubble rate becomes $H=\frac{m}{3t}$.

Geometrical analysis of dark energy models are usually performed through the statefinder pair $j$ and $s$ given by $j=\frac{\dddot{a}}{aH^3}$ and $s=\frac{j-1}{3(q-\frac{1}{2})}$. In the present model, we obtain these parameters as $j=\frac{9}{m}\left(\frac{2}{m}-1\right)+1$ and $s=\frac{2}{m}$. These parameters are constants of cosmic time and depend only on the exponent $m$. It can be emphaiszed here that, exact determination of these parameters from different observational bounds can constrain the exponent $m$ in narrow ranges.
\section{Dynamics of Anisotropic $BVI_h$ Universe}
Once the cosmic expansion behaviour is known, it becomes simpler to study the background cosmology of the diagonal $BVI_h$ universe. However, the exponent $h$ in the metric can assume integral values such as $-1, \ 0$ and $1$. Each value of $h$ corresponds to a different cosmological model with different dynamical behaviour. In view of this, in the following, we discuss the dynamical features of the three possible models in the frame work of $f(R,T)$ theory.

\subsection{Model-I ($h=-1$)} 
A substitution of $h=-1$ in eq. $\eqref{eq:14}$ yields $H_y=H_z$, where the integration constant has been rescaled to unity. Assuming an anisotropic relationship $H_x=kH_y$, we can write the functionals $\chi(H_x,H_y)$ and $\xi(H_x,H_y,H_z,h)$ as
\begin{eqnarray}
\chi &=& (k+1)\dot{H_y}+(k^2+k+1)H_y^2-\frac{1}{A^2},\label{eq:22}\\
\xi &=& (2k+1)H_y^2-\frac{1}{A^2},\label{eq:23}
\end{eqnarray}
where $k$ is an arbitrary positive constant. For a power law cosmology, we have $H_x=\left(\frac{km}{k+2}\right)\frac{1}{t}$, $H_y=H_z=\left(\frac{m}{k+2}\right)\frac{1}{t}$.  Consequently the directional scale factors are $A=t^{\frac{km}{k+2}}$ and $B=C=t^{\frac{m}{k+2}}$. Thus we can have
\begin{eqnarray}
\chi &=& \left[\frac{m^2(k^2+k+1)-m(k+1)(k+2)}{(k+2)^2}\right]\frac{1}{t^2}-\frac{1}{t^{\frac{2mk}{k+2}}},\label{eq:24}\\
\xi &=& \left[\frac{(2k+1)m^2}{(k+2)^2}\right]\frac{1}{t^2}-\frac{1}{t^{\frac{2mk}{k+2}}}.\label{eq:25}
\end{eqnarray}
The dynamical behaviour of the model is decided from the behaviour of the equation of state parameter $\omega$ and the effective cosmological constant $\Lambda$. However, these two parameters depend on the time varying nature of functionals $\chi(t)$ and $\xi(t)$ which in turn depend on the parameters $m$ and $k$. If $mk>k+2$, the terms within the square brackets in eqs. \eqref{eq:24} and \eqref{eq:25} dominate at late times of cosmic evolution whereas the terms containing $t^{-\frac{2mk}{k+2}}$ dominate at early phase of cosmic evolution. Here we wish to adopt a dimensional analysis to get some idea into the general behaviour of these functionals. Since $m$ and $k$ are two dimensionless constants, it appears that the dimensionality of the time dependent factors for a given functional should remain the same. In other words, we can have $m=1+\frac{2}{k}$ so that $\chi(t)$ and $\xi(t)$ become
\begin{eqnarray}
\chi(t) &=& \left(\frac{1-k^2}{k^2}\right)\frac{1}{t^2},\label{eq:26}\\
\xi(t) &=&\left(\frac{1+2k-k^2}{k^2}\right)\frac{1}{t^2}.\label{eq:27}
\end{eqnarray}

 \begin{figure}[h!]
  \includegraphics[width=\textwidth]{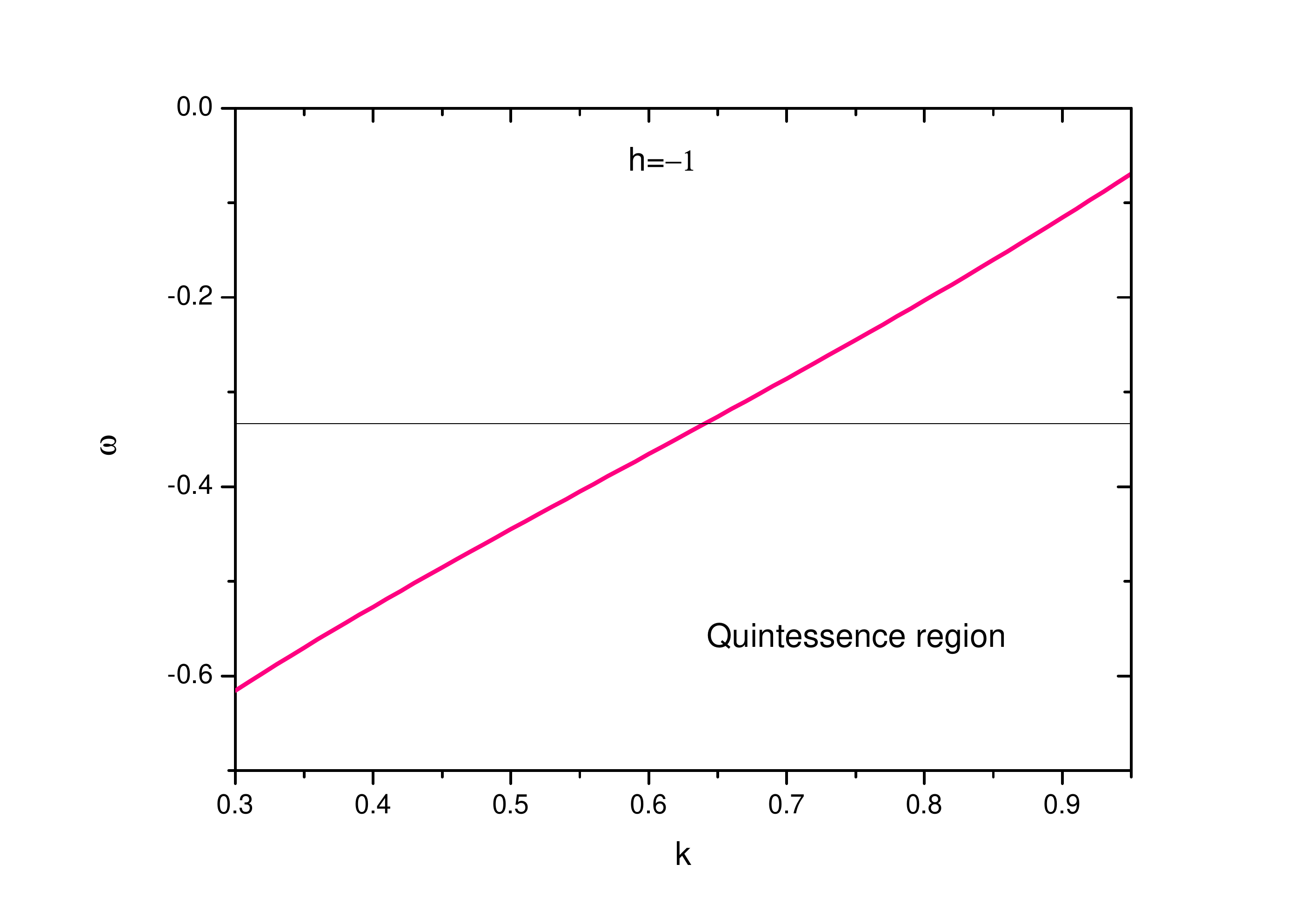}
  \caption{Variation of the equation of state parameter $\omega$ with the parameter $k$ for $h=-1$.}
  \label{fig1}
\end{figure}

The equation of state parameter and the effective time varying cosmological constant are obtained from \eqref{eq:17} and \eqref{eq:18} as
\begin{eqnarray}
\omega &=& 2\alpha+(4\alpha^2-1)\left[\frac{1+2k-k^2}{(1-2\alpha)(1-k^2)+4\alpha k}\right],\label{eq:28}\\
\Lambda (t) &=& \frac{2}{2\alpha +1}\left[\frac{k^2-k-1}{k^2}\right]\frac{1}{t^2}.\label{eq:29}
\end{eqnarray}

\begin{center}
\begin{figure}[h!]
  \includegraphics[width=\textwidth]{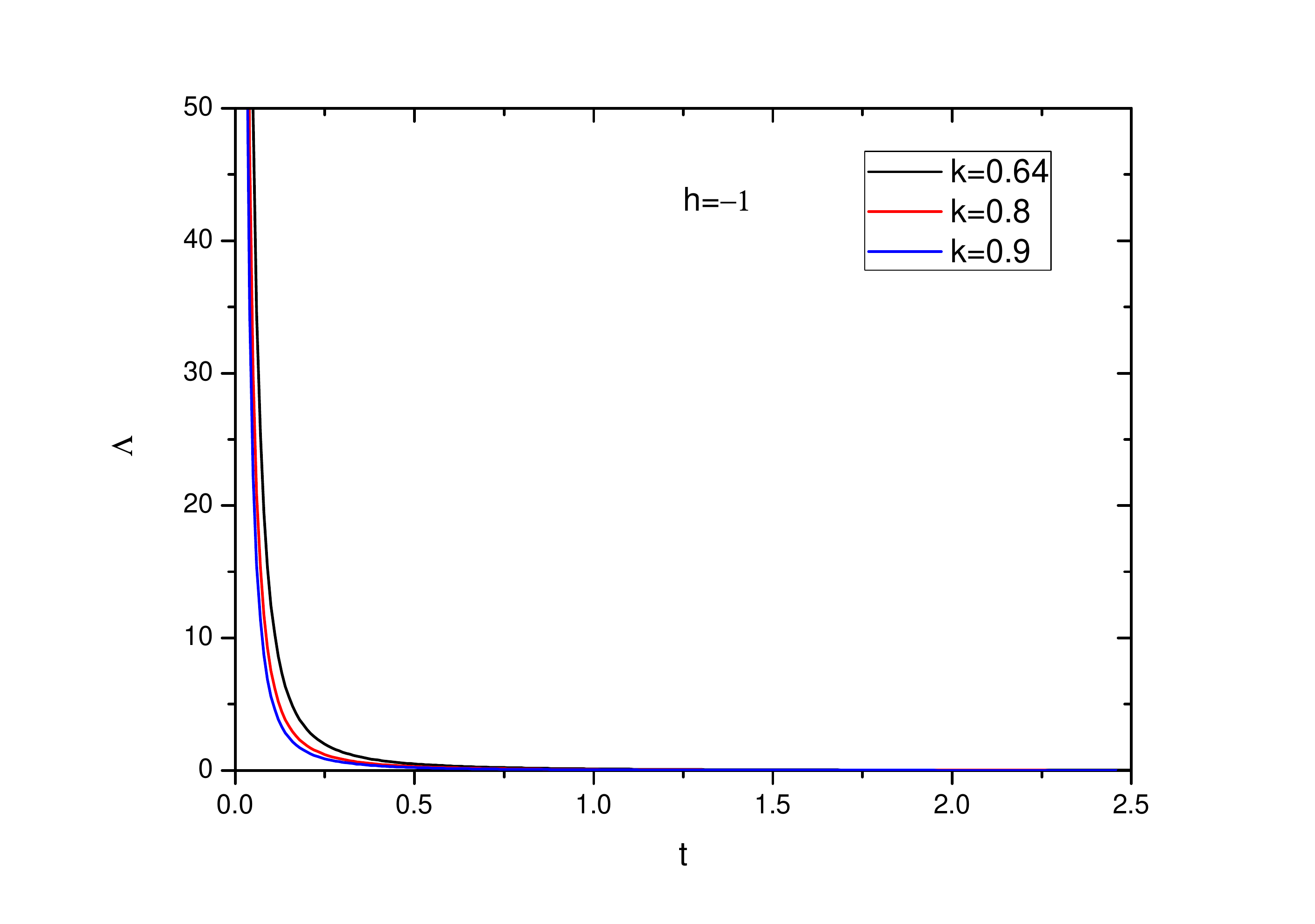}
  \caption{Evolution of the effective cosmological constant for three representative values of $k$ in the model $h=-1$.}
  \label{fig2}
\end{figure}
\end{center}
 
 It is obvious from the above expressions that, the equation of state parameter $\omega$ is a constant quantity for a given value of scaling constant $\lambda$ and the anisotropic parameter $k$. However, the effective cosmological constant decreases quadratically with cosmic time. In order to get viable cosmological models in conformity to recent observations, the cosmological constant should be dynamically varying from large positive values at an initial epoch to vanishingly null values at late times of cosmic evolution. Similarly, the equation of state parameter should be negative with values less than $-\frac{1}{3}$ at late times. This behaviour will enable us to constrain the parameter $k$. In Figure-1, we have shown the variation of $\omega$ as a function of $k$. Here we chose a small negative value of the scaling constant $\lambda$. The equation of state parameter $\omega$ increases almost linearly from a negative value for lower $k$ to zero at higher $k$. One can note that, the present model will collapse at $k=1$ and therefore, we restrict the values of $k$ below 1. For $k \leq 0.64$, $\omega$ remains in the quintessence region. In Figure-2, we have shown the dynamical variation of the effective cosmological constant for some representative values of $k$. As is required for an explanation to the late time cosmic acceleration, $\Lambda$ varies from a large positive values in the beginning to vanishingly small values at late times. As it appears from the figure, the behaviour of $\Lambda$ is less affected by the choice of the $k$ both at remote past and the future. However, the choice of $k$ affects $\Lambda$ at some epochs in recent past. In fact in these cosmic period, $\Lambda$ decreases with $k$ and the curves of $\Lambda$ move to the lower side for higher values of $k$.

\subsection{Model-II ($h=0$)}
In the present case with $h=0$, eq.\eqref{eq:14} reduces to $H_x=H_y$. An anisotropic relation $H_z=nH_y$ among the respective directional Hubble rates in the power law expansion of volume scale factor yields $H_x=H_y=\left(\frac{m}{n+2}\right)\frac{1}{t}$ and $H_z=\left(\frac{mn}{n+2}\right)\frac{1}{t}$. The directional scale factors become $A=B=t^{\frac{m}{n+2}}$ and $C=t^{\frac{mn}{n+2}}$. Here $n$ is a constant parameter. If $n=1$, the model reduces to be isotropic.

The functionals $\chi(t)$ and $\xi(t)$ for this model are obtained as
\begin{eqnarray}
\chi(t) &=& \left[\frac{3m^2-2m(n+2)}{(n+2)^2}\right]\frac{1}{t^2}-\frac{1}{t^{\frac{2m}{n+2}}}, \label{eq:30}\\
\xi (t) &=& \left[\frac{(2n+1)m^2}{(n+2)^2}\right]\frac{1}{t^2}-\frac{1}{t^{\frac{2m}{n+2}}}.\label{Eq:31}
\end{eqnarray}

The dimensional consistency of terms involved in the expressions of $\chi (t)$ and $\xi (t)$ constrains the exponent $m$ to be $m=n+2$. With this constraint, the equation of state parameter and the effective cosmological constant become
\begin{eqnarray}
\omega &=& \frac{1}{2\alpha}, \label{eq:32}\\
\Lambda (t) &=& -\frac{2n}{(2\alpha+1)t^2}. \label{eq:33}
\end{eqnarray}

 \begin{figure}[h!]
  \includegraphics[width=\linewidth]{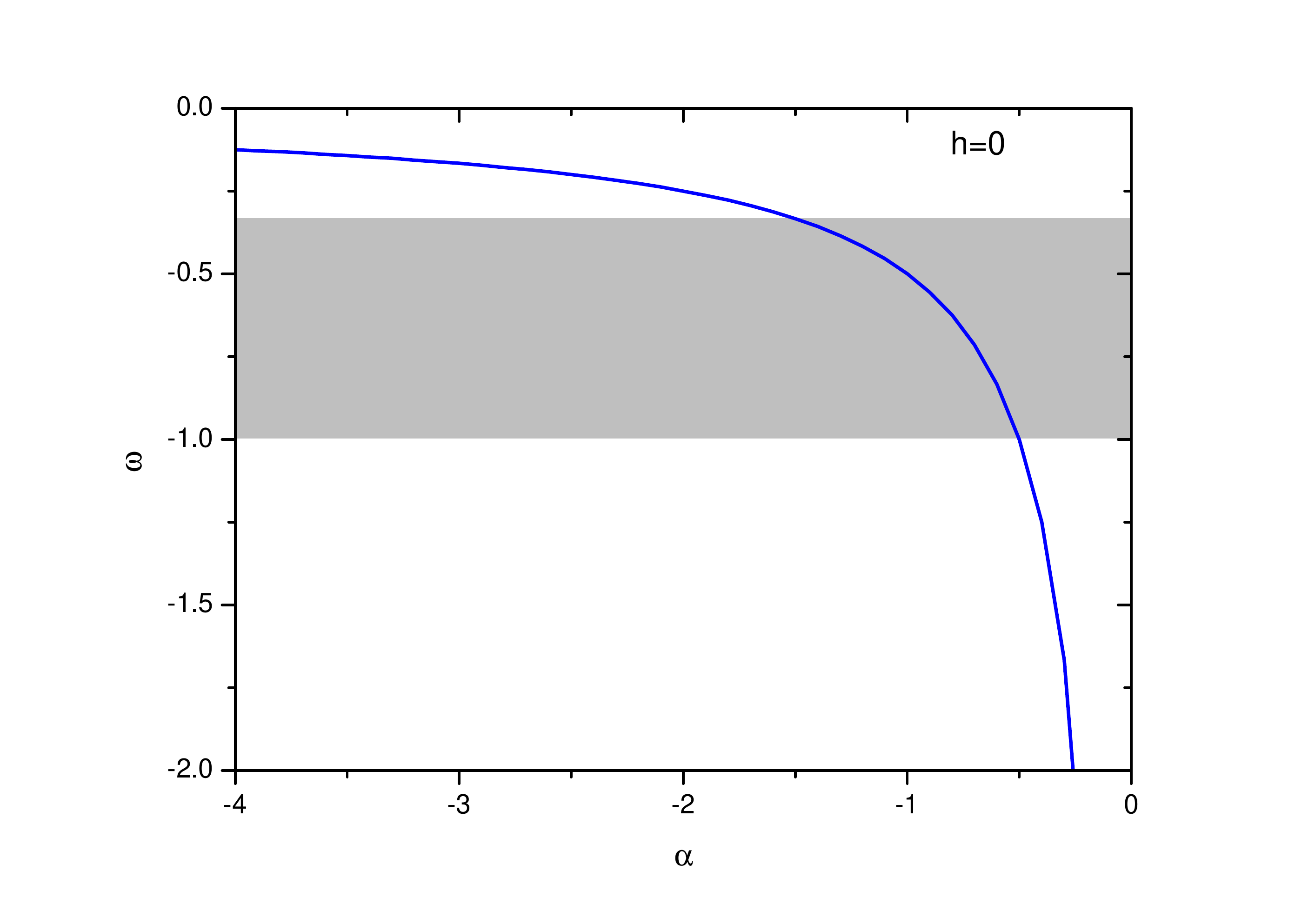}
  \caption{Variation of the equation of state parameter $\omega$ with the parameter $\alpha$ for $h=0$. The shaded portion denotes the quintessence region.}
  \label{fig3}
\end{figure}
\begin{figure}[h!]
  \includegraphics[width=\textwidth]{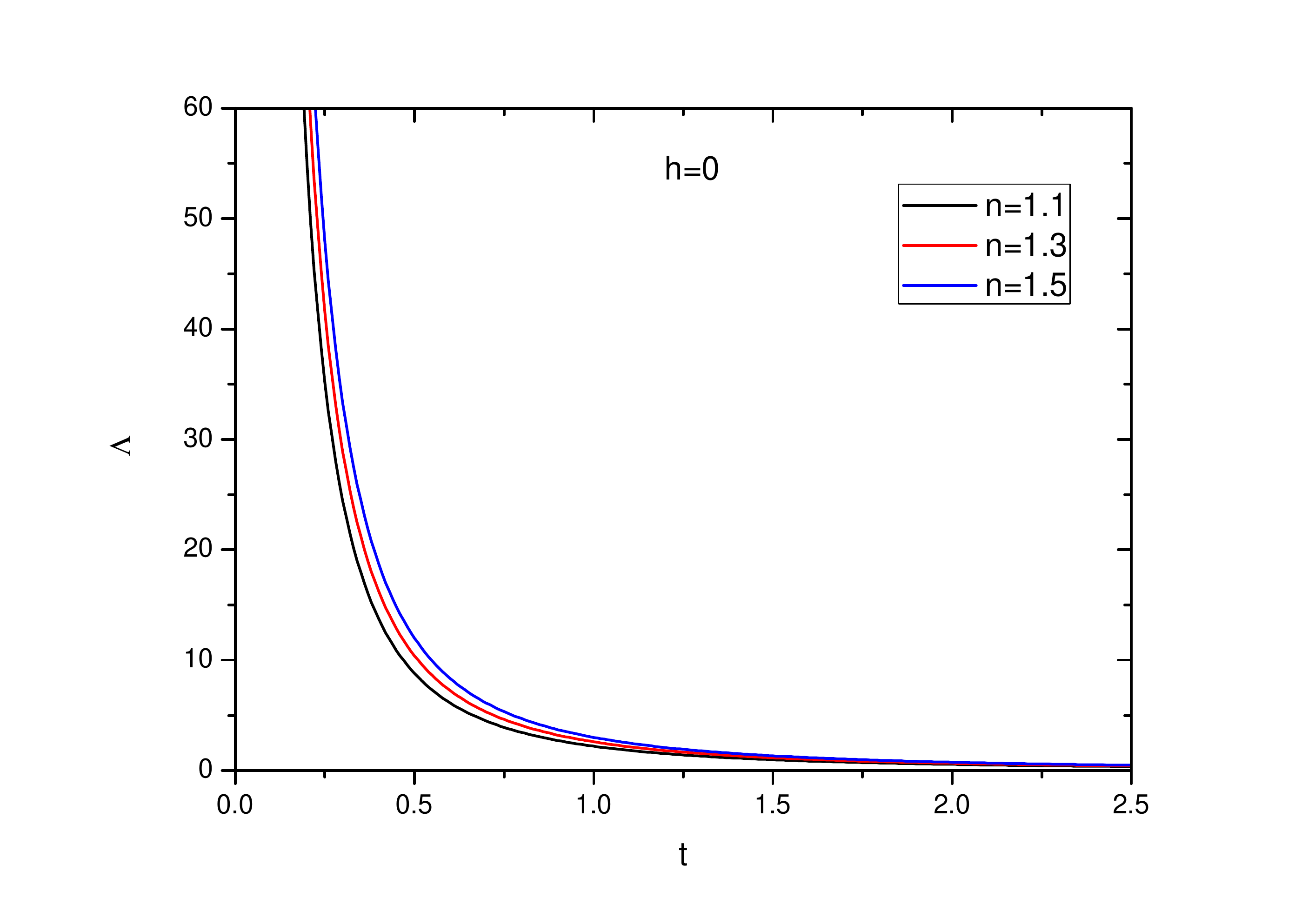}
  \caption{Evolution of the effective cosmological constant for three representative values of $n$ in the model $h=0$.}
  \label{fig4}
\end{figure}

As in the previous model, in this model also, the equation of state parameter is a constant quantity that depends on the scaling constant $\lambda$ through $\alpha$. In the present work, we chose $\lambda$ to assume a negative value so that $\alpha$ becomes negative. This puts $\omega$ in the negative domain. In Figure-3, we have plotted $\omega$ as a function of $\alpha$ in its negative domain. It is clear that, $\omega$ lies in the quintessence region (shaded portion in the plot) for the range $-1.5 \leq \alpha \leq -0.5$. For $\alpha >-0.5$, the equation of state parameter enters into the phantom region. It is certain from \eqref{eq:33} that the effective cosmological constant can be positive for $\alpha > -0.5$. In other words, an accelerated expansion with positive cosmological constant in this model favours a phantom phase. The time evolution of the effective cosmological constant is shown for different values of the anisotropic parameter $n$ in Figure-4. We have chosen the values of $n$ so as to get negative deceleration parameter. In order to satisfy this condition, $n$ has to be constrained in the range $n>1$. It is clear from the figure that, $\Lambda$ decreases from large positive values to small positive values during the cosmic evolution and vanishes at late times. One interesting thing in the present model is that, even if the model favours a phantom phase, the decrement in $\Lambda$ is bit slower than that of the previous model with $h=-1$.
\subsection{Model-III ($h=1$)}
In this model with $h=1$, we obtain from eqs. \eqref{eq:10} and \eqref{eq:11}
\begin{equation}
\frac{\dot{H}_x-\dot{H}_y}{H_x-H_y}+\theta=0,\label{eq:34}
\end{equation}
which can be integrated for the power law cosmology to get
\begin{equation}
H_x=H_y+\frac{\epsilon}{t^m}.\label{eq:35}
\end{equation}
Here the integration constant $\epsilon$ is related to the present day value of the directional Hubble parmeters as $\epsilon=H_{x0}-H_{y0}$. 

Eq. \eqref{eq:14} becomes $2H_x=H_y+H_z$ which implies $H_x=H$ and consequently
\begin{equation}
H_y=H-\frac{\epsilon}{t^m},  ~~~~~ H_z=H+\frac{\epsilon}{t^m}.\label{eq:36}
\end{equation}
One can note that, since the dimension of $H$ is that of $t^{-1}$, $\epsilon$ has a dimension of $t^{m-1}$.

The functionals $\chi (t)$ and $\xi (t)$ are obtained as
\begin{eqnarray}
\chi(t) &=& 2\dot{H}+3H^2+\frac{\epsilon^2}{t^{2m}}-\frac{1}{t^{\frac{2m}{3}}},\\ \label{eq:37}
\xi (t) &=& 3H^2-\frac{\epsilon^2}{t^{2m}}-\frac{3}{t^{\frac{2m}{3}}}.\label{eq:38}
\end{eqnarray}

From dimensional consistency, the parameter $m$ can be constrained as $m=3$. In order to get accelerating models, the deceleration parameter $q$ should be negative which requires that $m$ should be greater than 3. However, the dimensional analysis yields $q=0$ for the present model. Also, interestingly the equation of state parameter and the effective cosmological constant are obtained to be $\omega=1$ and $\Lambda=0$. Even though, a vanishing cosmological constant is acceptable, $\omega=1$ may not be acceptable in the context of dark energy driven cosmic acceleration. In view of this, the $BVI_1$ model may not be in conformity with the present day observations.

\section{Conclusion}
In view of the recent interest in modified theories of gravity, we have studied the dynamics of some Bianchi type $VI_h$ models in $f(R,T)$ theory. We choose $f(R,T)= f(R)+f(T)$ where $f(R)=\lambda R$ and $f(T)=\lambda T$. These linear functions $f(R)$ and $f(T)$ rescale the modified gravity theory and generates the concept of a time varying effective cosmological constant. We have investigated three different models corresponding to three values of the metric parameter $h$ i.e. $-1, 0$ and $1$. The dynamics of the models are studied for a presumed power law expansion of the volume scale factor. We have adopted dimensional analysis method to constrain some of the model parameters. In the anisotropic models with $h=-1$ and $h=0$, the effective cosmological constant is found to evolve from large positive values at the beginning to small values at late times. This result is in accordance with the observations concerning the dark energy driven cosmic acceleration. The equation of state parameter for these two models becomes negative. While in the first model, it remains within the quintessence region for some acceptable value of the anisotropic parameter $k$, for the second model, it enters into the phantom region. However, for the third model with $h=1$, viable cosmological models could not be obtained.
\section{Acknowledgement}
BM acknowledges SERB-DST, New Delhi, India for financial support to carry out the Research project[No. SR/S4/MS:815/13]. SKT and BM thank IUCAA, Pune for the support in an academic visit to accomplish a part of this work.


\begin{thebibliography}{99}

\bibitem{Reiss98} A.G. Riess et al.,  \textit{Astronomical Journal}, vol. 116, no. 3, pp. 1009-1038.. 1998..
\bibitem{Perlm99} S.J. Perlmutter  et al., \textit{The Astrophysical Journal}, vol. 517, no. 2, pp. 565-586, 1999.
\bibitem{Bahcall99} N.A. Bahcall, J.P. Ostriker,  S. Perlmutter and P.J. Steinhardt,\textit{Science}, vol. 284 no. 5419, pp.  	1481-1488, 1999.
\bibitem{Tegmark04} M. Tegmark et al., \textit{Physical Review D}, vol. 69, Article ID 103501, 2004.
\bibitem{Abaz04} K. Abazajian, \textit{The Astronomical Journal}, vol. 128, no. 1, pp. 502-512, 2004.
\bibitem{Pope04} A.C.  Pope et al., \textit{The Astrophysical Journal}, vol 607, no. 2, pp. 655-660, 2004.
\bibitem{Bennet03} C.L. Bennett et al.,\textit{Astrophys. J. Suppl}, vol. 148, pp. 1-27,  2003.
\bibitem{Spergel03} D.N. Spergel et al., \textit{Astrophys. J. Suppl}, vol. 148, pp. 175-194,  2003.
\bibitem{Nojiri06} S.Nojiri and S.D. Odintsov, \textit{Physical Review D}, vol. 74, Article ID. 086005, 2006.
\bibitem{Nojiri07} S.Nojiri and S.D. Odintsov, \textit{Int. J. Geom. Methods Mod. Phys.}, vol. 04, no. 1, 115-145, 2007.
\bibitem{Linder10} E.V. Linder, \textit{Physical Review D}, vol. 82, Article ID 109902, 2010.
\bibitem{Myrz11} R. Myrzakulov, \textit{Eur. Phys. J. C}, vol. 71, Article ID, 1752, 2011.
\bibitem{Chen11} S.H. Chen, J. B. Dent, S. Dutta and E. N. Saridakis, \textit{Physical Review D}, vol. 83, Article ID 023508, 2011.
\bibitem{Dent11} J.B. Dent, S. Dutta and E. N. Saridakis, \textit{JCAP}, vol. 2011, Article ID 009, 2011. 
\bibitem{Harko14} T. Harko, F. S. N. Lobo, G. Otalora and E. N. Saridakis, \textit{Phys. Rev. D}, vol. 89, Article ID 124036, 2014.
\bibitem{Nojiri05} S.Nojiri and S.D. Odintsov, \textit{Phys Lett B}. vol. 631, no. 1-2, PP. 1-6, 2005.
\bibitem{Li07} B. Li et al., \textit{Phys Rev D}, vol. 76, Article iD 044027, 2007.
\bibitem{Kofinas14} G. Kofinas and E.N. Saridakis, \textit{Phys. Rev. D}, vol. 90, Article ID 084044, 2014.
\bibitem{Harko11} T. Harko, F.S.N. Lobo, S. Nojiri and S.D. Odintsov, . \textit{ Phys. Rev. D}, vol. 84, Article ID 024020, 2011.
\bibitem{Myrz12} R. Myrzakulov, \textit{Eur. Phys. J. C}, vol. 72, Article ID 2203, 2012.
\bibitem{Sharif12} M. Sharif and M. Zubair J. \textit{Phys. Soc. Jpn.}, vol.  81, Article ID 114005, 2012.
\bibitem{Houndjo12} M. J. S. Houndjo and O. F. Piattella, \textit{Int. J. Mod. Phys. D}, vol.  21, Article ID 1250024, 2012.
\bibitem{Alva13} F. G. Alvarenga, M. J. S. Houndjo, A. V. Monwanou, and J. B. Chabi Orou, \textit{J. Mod. Phys.} vol. 04, pp. 130, 2013.  
\bibitem{Jamil12} M. Jamil et al., \textit{Eur. Phys. J. C}, vol.  72, pp. 1999, 2012.
\bibitem{Shamir12} M. F. Shamir, A. Jhangeer and A. A. Bhatti, \textit{arXiv:1207.0708v1} [gr-qc], 2012.
\bibitem{Chau13} R Chaubey and A K Shukla, \textit{Astrophys. Space Sci.}, 343, 415, 2013.
\bibitem{Ahmed14} N Ahmed and A Pradhan, \textit{Int. J. Theor. Phys.}, 53 289, 2014.
\bibitem{Yadav13} A K Yadav, \textit{arXiv:1311.5885v1}, 2013. 
\bibitem{Singh14} C.P. Singh, P. Kumar, Eur. Phys. J. C 74, 3070, 2014.
\bibitem{Mishra14} B.Mishra, P.K. Sahoo, \textit{Astrophys. Space Sci.}, 352(1), 331-336, 2014.
\bibitem{Samanta13} G.C. Samanta, \textit{Int. J. Theo. Phys.}, vol. 52, pp. 2647-2656, 2013.
\bibitem{Shamir15} M.F. Shamir, \textit{Int. J. Theor. Phys.}, 54, 1304, 2015. 
\bibitem{Mishra15} B. Mishra, P.K. Sahoo, S. Tarai, \textit{ Astrophys Space Sci.}, 359, Article ID 15, 2015.
\bibitem{Sahoo16} P. K. Sahoo, B. Mishra, S. K.Tripathy, \textit{ Indian Jou. Phys.}, Vol.90, 485-493, 2016.


\end{thebibliography}
\end{document}